\documentstyle[preprint,aps,epsf]{revtex}
\begin{document}
\preprint{CPT-97/P.3456}
\title{Manifestation of a strong electroweak sector with decoupling at 
hadronic colliders}
\author{R. Casalbuoni and D. Dominici}
\address{Dipart. di Fisica, Univ. di Firenze, Largo E. Fermi, 2, I-50125
Firenze\\
and I.N.F.N., Sezione di Firenze, Largo E. Fermi, 2, I-50125 Firenze}
\author{P. Chiappetta and A. Deandrea}
\address{Centre de Physique Th\'eorique\footnote{Unit\'e Propre de 
Recherche 7061}, CNRS Luminy, Case 907, F-13288 Marseille Cedex 9}
\author{S. De Curtis}
\address{I.N.F.N., Sezione di Firenze, Largo E. Fermi, 2, I-50125 Firenze}
\author{R. Gatto}
\address{D\'epartement de Physique Th\'eorique, Univ. de Gen\`eve, 
24 quai E.-Ansermet, CH-1211 Gen\`eve 4}

\date{February 1997}
\maketitle

\begin{abstract}
We study the expected phenomenology at present (Tevatron) and future (Tevatron 
Upgrade, LHC) hadronic colliders of a model describing 
a strong electroweak symmetry breaking sector with both vector and axial 
vector strong interacting bosons degenerate in mass.
Due to decoupling, this model at low energies is almost indistinguishable 
from the Standard Model, passing therefore all low energy precision tests 
at LEPI. We will show that it gives quite visible signals at  
forthcoming hadronic accelerators. The new charged vector bosons can be
detected more easily than the neutral ones.
\end{abstract}

\section{Introduction}

Future hadron colliders offer the possibility of testing the
electroweak symmetry breaking sector by studying $WW$ scattering. It
is well known that if no new physics, like supersymmetry,
enters to guarantee a low mass Higgs, some new strong interaction must
become manifest at a scale of the order of 1 TeV. The standard
technicolor \cite{techni} or the walking technicolor \cite{walking}
are particular models of a dynamical breaking of the electroweak
symmetry. Their rich phenomenology at hadron colliders has been
recently reexamined \cite{lane}. One can also perform a model
independent analysis of a strong interacting electroweak sector, by
studying chiral effective lagrangians. Usually one takes into account
the following three possible scenarios. In the first no resonance in
the $WW$ channel is formed and one has just the sector of the
longitudinal  gauge bosons strongly interacting \cite{appel}. In the
second scenario the $WW$ interactions give rise to a scalar resonance
(Higgs remnant) \cite{einhorn}. In the third scenario the resonance is
formed in the $J=I=1$ channel, and a new triplet of gauge vector
bosons is present \cite{bess}. The phenomenology at hadron colliders
of these three different scenarios has been largely studied \cite{lhc1}
and it has been recently reviewed in ref. \cite{han}.

LEPI results have drastically constrained the strong mechanism for electroweak
symmetry breaking, namely rescaled QCD technicolor 
scenarios are ruled out. In this
paper we will focus on a scheme of dynamical symmetry breaking in which 
decoupling is 
automatically satisfied in the low energy limit, therefore the
deviations
from the Standard Model (SM) are strongly suppressed.
 This model, called degenerate BESS (BESS standing for Breaking 
Electroweak Symmetry Strongly), has been proposed in a previous work 
\cite{debess}, to which we refer the reader for details. 

In brief the spirit of the model is the following. We start from the global 
symmetry group of the theory $G=SU(2)_L \otimes SU(2)_R$, spontaneously broken
to the diagonal subgroup $H_D=SU(2)_V$. The new vector and axial vector 
bosons correspond to the gauge bosons associated to a hidden symmetry,
$H'=SU(2)_L \otimes SU(2)_R$. The symmetry group of the theory becomes
$G'=G\otimes H'$. It breaks down spontaneously to $H_D$ and gives rise to
nine Goldstones. Six of these are absorbed by the vector and
axial vector bosons. As soon as we perform
the gauging of the subgroup $SU(2)_L \otimes U(1)_Y \subset G$, 
the three remaining Goldstones disappear giving masses to the 
SM gauge bosons. This general procedure
for building models with vector and axial vector
resonances  is discussed in \cite{assiali}. 

What makes the model \cite{debess} so attractive is the fact that all the
deviations in the low 
energy parameters from their SM values are strongly 
suppressed. This allows the existence of a strong electroweak
sector
 at 
relatively low energies within the precision of electroweak tests, 
such that it may be accessible with accelerators designed for the near future.
As such it offers 
possibilities of experimental tests even with future or existing
machines of relatively low energy. The phenomenological implications will be 
discussed below.  

The study of the virtual effects of the heavy resonances has previously 
been done, so we shall now concentrate on their direct production at 
hadronic colliders. The Tevatron limits
on $W'$ are used to constrain the parameter space of degenerate BESS. A 
feature of degenerate BESS, as compared to BESS with only vector resonances, 
is the absence of direct coupling of the new resonances to the 
longitudinal weak gauge bosons. This implies larger widths into
fermion pairs as compared to widths into pairs of weak gauge bosons. 

Our phenomenological applications include discussion of the properties of the
heavy resonances (masses, partial widths) and studies of their effects
at the
Tevatron upgrade and at the LHC, including rough detector simulation. 

In section 2 we recall briefly the main features of the model and give 
the relevant formulas for the evaluation of the heavy resonances 
direct production. Subsequently we study the phenomenology related to 
their discovery at hadron colliders.

\section{The degenerate BESS model}

The model includes two new triplets of vector bosons ($L^\pm$, $L_3$, $R^\pm$,
$R_3$). The parameters of 
the model are a new gauge coupling constant $g''$ and a mass parameter $M$, 
related to the scale of the underlying symmetry breaking sector.
We therefore parameterize the model in terms of ${g''}$ and $M$. In the
following we give approximate formulas in the limit $M \to \infty$ and
$g'' \to \infty$ \cite{rob}. For the numerical analysis the exact
formulae of \cite{debess} were used.

In the charged sector the fields $R^\pm$ are unmixed for any value of ${g''}$.
Their mass is given by:
\begin{equation}
M^2_{R^\pm} \equiv M^2
\label{8.2}
\end{equation}

The charged fields $W^\pm$ and $L^\pm$ have the following masses:
\begin{eqnarray}
M^2_{{W}^\pm}&=&\frac{v^2}{4} { g}^2
,\nonumber\\
M^2_{{L}^\pm}&=&M^2 (1+2 {x^2})
\label{8.5}
\end{eqnarray}
where $x=g/g''$,  $g$ is the usual $SU(2)$
gauge coupling constant and $v^2=1/({\sqrt{2} G_F})$.

In the neutral sector we have:
\begin{eqnarray}
M^2_{Z}&=&\frac{M^2_W}{c^2_{\theta}}\nonumber\\
M^2_{L_3}&=& M^2\left(1+2 x^2\right)\nonumber\\
M^2_{R_3}&=& M^2\left(1+2 x^2 \tan^2 \theta\right)
\label{8.16}
\end{eqnarray}
where $\tan \theta = s_{\theta}/c_{\theta} = g'/g$ and $g'$ is the usual 
$U(1)_Y$ gauge 
coupling constant. Notice that for small $x$ all the new vector
resonances are degenerate in mass.

The charged part of the fermionic lagrangian is
\begin{equation}
{\cal L}_{charged}=
-\left(a_W W_\mu^-+a_L L_\mu^-\right)J_L^{(+)\mu}+ H.c.
\end{equation}
where
\begin{eqnarray}
a_W&=&\frac {g}{\sqrt{2}}\nonumber\\
a_L&=& -g x
\label{9.2}
\end{eqnarray}
apart from higher order terms and $J_L^{(+)\mu}={\bar \psi}_L \gamma^{\mu}
\tau^+ \psi_L$ with $\tau^+$ the combination
$(\tau_1 + i\tau_2)/2$. Let us notice that the $R^{\pm}$ are not coupled 
to the fermions.

For the neutral part we get
\begin{eqnarray}
{\cal L}_{neutral}&=& -\Big\{ eJ_{em}^\mu\gamma_\mu +
\left[A J_L^{(3)\mu}+BJ_{em}^\mu \right]Z_\mu\nonumber\\
&+&\left[C J_L^{(3)\mu}+DJ_{em}^\mu \right]L_{3\mu}\nonumber\\
&+&\left[E J_L^{(3)\mu}+FJ_{em}^\mu \right]R_{3\mu}\Big\}
\end{eqnarray}
where $\gamma_\mu$ in the preceeding formula is the photon field and 
again in the limit $M\to\infty$, $x\to 0$,
\begin{eqnarray}
&A&= \frac{g}{c_{\theta}}
~~~~B= -\frac{g s^2_\theta}{c_\theta} \nonumber \\
&C&=-\sqrt{2} g x ~~~~D= 0\nonumber \\
&E&= \sqrt{2} g \frac{x}{c_\theta} \tan^2 \theta ~~~~F= -E
\end{eqnarray}
and $J_{em}^{\mu}= Q {\bar \psi} \gamma^{\mu} \psi$, 
$J_L^{(3)\mu}={\bar \psi}_L \gamma^{\mu} T_L^3 \psi_L$ are the usual neutral
currents.

The total fermionic widths are
\begin{eqnarray}
\Gamma^{fermion}_{L_3}&=&\frac{2\sqrt{2}G_FM_W^2}{\pi}M_{L_3}
\left(\frac g {g''}\right)^2\nonumber\\
\Gamma^{fermion}_{R_3}&=&\frac{10\sqrt{2}G_FM_W^2}{3\pi}\frac {{s_\theta}^4}
{{c_\theta}^4}M_{R_3}\left(\frac g {g''}\right)^2\nonumber\\
\Gamma^{fermion}_{L^\pm}&=&\frac{2\sqrt{2}G_FM_W^2}{\pi}M_{L^\pm}
\left(\frac g {g''}\right)^2
\end{eqnarray}
while from the trilinear gauge couplings we get the following widths:
\begin{eqnarray}
\Gamma^{WW}_{L_3}&=&\frac{\sqrt{2}G_FM_W^2}{24\pi}M_{L_3}
\left(\frac g {g''}\right)^2\nonumber\\
\Gamma^{WW}_{R_3}&=&\frac{\sqrt{2}G_FM_W^2}{24\pi}
\frac{{s_\theta}^4}{{c_\theta}^4}
M_{R_3}\left(\frac g {g''}\right)^2\nonumber\\
\Gamma^{WZ}_{L^\pm}&=&\frac{\sqrt{2}G_FM_W^2}{24\pi}M_{L^\pm}
\left(\frac g {g''}\right)^2
\end{eqnarray}
It may be useful to compare the widths of the new gauge bosons into
vector boson pairs with those into fermions:
\begin{eqnarray}
\Gamma_{L_3}^{fermion}&=&48~\Gamma_{L_3}^{WW} \nonumber\\
\Gamma_{R_3}^{fermion}&=&80~\Gamma_{R_3}^{WW} \nonumber\\
\Gamma_{L^\pm}^{fermion}&=&48~\Gamma_{L^\pm}^{WZ}
\end{eqnarray}
We see that the total fermionic channel is dominant due to the multiplicity.

\section{Degenerate BESS at Tevatron}

Data from the Fermilab Tevatron Collider, collected by the CDF collaboration
\cite{cdf} put limits on the model parameter space. Their search was done
through the decay $W' \to e \nu_e$, assuming standard couplings of the $W'$ to
the fermions. Their result can be translated into a limit for the 
degenerate BESS model parameter space. 

In Fig. 1 these limits are
shown in terms of the mass of the $R^\pm$ resonance, $M$, 
and the ratio of coupling constants, $x=g/g''$. 
The present limit from CDF obtained with an integrated luminosity of 19.7 
$pb^{-1}$ (dashed line) is compared with the result obtained from
LEPI (continuous line). Limits from LEPI are obtained calculating virtual
effects up to order $M_W^2/M^2$, and using the experimental data from ref.
\cite{altarelli}. Note that in the low energy limit 
($M \to \infty$) there are no deviations from the SM, thus 
allowing to consider light new resonances for the strong sector. 
The figure was obtained using the CDF 95\% C.L. limit
on the $W'$ cross-section times the branching ratio and comparing this limit 
with the predictions of our model at fixed $g/g''$, thus 
giving a limit for the $R^\pm$
mass. This procedure was then iterated for various values of $g/g''$. 
The statistical significance of the plot is that of a 95\% C.L. limit in one
variable, the mass, at a given value of $g/g''$. The CDF limit is presently 
less restrictive than the LEPI one.
While waiting for the analysis of  100 $pb^{-1}$, we have considered
an extrapolation of the CDF data based on the principle that when the
background is present the cross-section limit scales inversely with
the square root of the luminosity \cite{TeV2000}. The extrapolation of 
the CDF bound to an integrated luminosity of 100 $pb^{-1}$ (dotted line)
improves marginally LEPI result and only for masses of the new vector 
resonances smaller than 400 GeV. 

We will now study the detection of charged and neutral vector resonances 
from a strong electroweak sector at the upgrading of the Fermilab
Tevatron~\cite{TeV2000}. The option we have chosen is the so called 
TeV-33, with 
a c.m. energy of the collider of 2 TeV and an integrated luminosity of 
$10~fb^{-1}$. 
We have considered the total 
 cross-section $p {\bar p} \to L^{\pm},W^{\pm}\to
\mu \nu_\mu$ and compared it with the SM background.
 A minimum of 10 events per year has been 
required to detect the signal with respect to the background.
If no deviations are seen within the 
statistical error and a systematic error of 5\% on the cross-section,
we get the  90\% C.L. bound shown in Fig. 2.

Up to a region of 1 TeV the limits from TeV-33
option are stronger with respect
to LEPI (we recall that LEPII will only marginally improve LEPI
results \cite{debess}).

The events where simulated using Pythia Montecarlo \cite{phy}. The 
simulation was performed using the expected detector resolution, in particular
a smearing of the energy of the leptons was done according to
$\Delta E/{\sqrt{E}} = 10\%$ and the error in the 3-momentum determination 
was assumed to be between 3\% for a mass of the order of 500 GeV and 5\% for 
a mass of 1000 GeV. 

We have first analyzed the production of the charged resonances 
$L^\pm$ of the degenerate BESS in the channel $p {\bar p} \to \mu \nu_\mu$. 
Only Drell-Yan mechanism for production was considered since it is the
dominant one (see the couplings and widths listed in the previous
section). 
The signal events were compared with the background from SM 
muon-neutrino production. Two cases have been analyzed, namely a mass 
of 400 GeV for the new charged resonances with 
$g/g''=0.12$ (Fig. 3) and $M=600$ GeV with $g/g''=0.2$ (Fig. 4). 
The first case is 
taken in a region of the parameter space, which is allowed at present, but 
is expected to be soon excluded at Tevatron when all data 
will be analyzed (100 
$pb^{-1}$). The second choice of the parameters of the model is instead
testable only with the TeV-33 option. The observable we deal with is the 
transverse mass of the charged resonance.

We now consider the production of the neutral resonances of the model, 
and their subsequent decay into muon pairs. This is the cleanest signature,
but the production rate is less favourable than for the new charged bosons.
As a consequence the best exclusion limits that can be imposed on the parameter 
space originate from non-observation of charged resonances.
In order to compare the neutral boson production to the charged one, we have
taken in Figs. 5 and 6, the same values for the parameters as in Figs. 3 and 4.
The observable in this case is the invariant mass of the muon pair.

We have examined various cases with different choices of $M$, and $g/g''$
(taken inside the physical region shown in Fig. 1)
to give an estimate of the sensitivity of the model to this
option for the upgrading  of the Tevatron (see Tables 1 and 2).
For each case we have selected cuts to maximize the statistical 
significance of the signal.
We see that the number of signal events decreases for increasing mass of the 
resonance. The conclusion is that Tevatron with the high luminosity option 
will be able to discover a strong electroweak resonant sector as 
described by the degenerate BESS model for masses up to 1 TeV. It can be 
seen from the calculation of the statistical significance (see Tables 1 and 2)
that the charged process allows to push further the discovery limits
of the new vector bosons at Tevatron with respect to the neutral
process.
However the experimental check of the model requires the proof of the
existence of both neutral and charged vector bosons.
Notice that the reconstruction of the resonance mass 
requires a careful study 
of the experimental setup, due to the smallness of the resonance width.
 
\section{Degenerate BESS at the LHC}    

The physics of the Large Hadron Collider (LHC) has been discussed in a
number of papers (see \cite{LHC} and references therein); it
will be able either to discover the new resonances or to constrain
the physical region left unconstrained by previous data.
We have considered a configuration of LHC with a c.m. energy $\sqrt{s}=14$ TeV,
a luminosity of $10^{34} cm^{-2} s^{-1}$  and one year run ($10^7$ s). 

If no new resonances are discovered, limits can be imposed on the parameter 
space of the model. We have  considered the total cross-section of 
$pp \to L^\pm, W^\pm \to \mu \nu_\mu$, which has a clear signature 
and a large number of events, and we have compared it with the SM 
background of $\mu \nu_\mu$. We have obtained a contour plot at 90\% C.L. 
in the two variables $M$ and $g/g''$, shown in Fig. 7.

The applied cuts were $\vert p_{t\mu}\vert > M/2 -50~ GeV$ in order 
to maximize the deviation of BESS model cross-section with 
respect to the SM one (most of the SM events are low $p_t$ ones). 
Moreover we have assumed a systematic 
error of 5\% in the cross-section and the statistical error obtained 
considering a luminosity of $10^{34} cm^{-2} s^{-1}$ (dotted-dashed line) 
or of $10^{33} cm^{-2} s^{-1}$ (dotted line) and one year run ($10^7$ s). 
The new resonances of the model can be discovered directly for a wide range 
of values of the parameter space of the model. The discovery limit in the 
mass of the resonance depends on the value of $g/g''$. 
For example if $g/g''=0.1$, the resonance is visible over the background at 
least up to 2 TeV. 

In Figs. 8-12 we show the differential distribution of events at LHC of 
$pp \to L^\pm, W^\pm \to \mu \nu_\mu$  in the transverse mass of the new vector
boson for different values of $M$ and $g/g''$. 
The choice we have done is within the region not accessible
to the upgrading of the Tevatron. The energy of the 
muons was smeared by 10\% and the error in the 3-momentum increases with the 
momentum of the muon from 3\% to 9\% as stated in \cite{atlas}.

In particular in Fig. 8 a light resonance mass $M=500~GeV$ and 
$g/g''=0.15$ is considered. Once cuts have been applied the number of signal 
events per year 
is approximately 85000 whereas the corresponding background consists of 26000 
events. The smearing of the muon momentum is taken equal to 3\% in this case.

In Fig. 9 only $g/g''$ is changed, in order to illustrate the LHC sensitivity 
to the BESS parameters. When $g''$ gets larger the number of events decreases
as $1/g''^2$.

In Figs. 10-12 we show cases corresponding respectively to 
$M=1, 1.5, 2$ TeV with $g/g''=0.1$. 
The smearing of the muon momentum is taken equal to 5, 7, 9\%
respectively.

The statistical significance of the previous examples is given in Table 3,
showing that the discovery of a charged resonance up to 2 TeV with $g/g''
\simeq 0.1$ is well within the reach of LHC. The limit of detection for a 
2 TeV mass is reached for a value of $g/g''=0.03$.

We will finally consider the production and decay 
of the corresponding neutral resonances of the model in Figs. 13-17, using
the same values for the BESS parameters as in the charged channel
examined before.

Although the number of events is smaller, neutral vector bosons up to 2 TeV,
provided $g/g''$ is not smaller than 0.1, can be seen (see Table 4). 

\section{Conclusions}

We have discussed an effective theory describing new vector and axial vector 
resonances within the scheme of a strong electroweak breaking sector.

The new vector and axial vector particles are degenerate in mass (at the 
leading order) and their virtual effects are suppressed, allowing the model
to pass all low energy precision tests at LEPI. In the low energy 
limit ($M\to \infty$ with the gauge coupling of the new resonances fixed) the 
new particles are decoupled and we obtain the effective lagrangian of the 
SM. 
High energy hadron colliders, as Tevatron and LHC, will allow to study the 
direct production of new resonances from Drell-Yan mechanism. With actual 
Tevatron luminosity, it is not possible to improve LEPI bounds on degenerate
BESS parameter space. This is no more the case when the upgrade in luminosity 
of Tevatron is considered. If the new particles are not discovered at LHC, 
stringent bounds on the  model parameters  are derived.

The direct observation of the new resonances is 
possible in a wide range of the parameter space of the model, up to the 
1 TeV range at Tevatron and up to 2 TeV at LHC, in some case with a 
spectacular number of events. 

\par\noindent
\vspace*{1cm}
\par\noindent
{\bf Acknowledgements}
\par\noindent
This work has been carried out within the Program Human Capital and
Mobility: ``Tests of electroweak symmetry breaking
and future European colliders'', CHRXCT94/0579, BBW/OFES 95.0200.
A.D. acknowledges the support of a TMR research fellowship of the European
Commission under contract nr. ERB4001GT955869. We thank D.Fouchez for
discussion on PYTHIA Montecarlo at an early stage of this work.

\begin{table}
\hfil
\vbox{\offinterlineskip
\halign{&#& \strut\quad#\hfil\quad\cr
\tableline
\tableline
&$g/g''$ && $M$ && $\Gamma_{L^\pm}$ && $\vert p_t^\mu\vert_c$ && 
$m_T$ &&  $\# B$ && $\# S$ && $S/{\sqrt{S+B}}$ & \cr
\tableline
&\omit&&GeV&&GeV&&GeV&&GeV&&\omit&&\omit&&\omit& \cr
\tableline
&\omit&&\omit&&\omit&&\omit&&\omit&&\omit&&\omit&&\omit&\cr
&0.12 && 400 && 0.4 && 150 && 300 && 385 && 887&& 24.9 & \cr  
&0.20 && 600 && 1.7 && 200 && 400 && 82 && 303 && 15.4 & \cr  
&0.40 && 1000 && 11.1 && 300 && 800 && 0 && 16 && 4.0 & \cr 
&\omit&&\omit&&\omit&&\omit&&\omit&&\omit&&\omit&&\omit&\cr
\tableline
\tableline}}
\caption{Degenerate BESS at Tevatron Upgrade for the process
 $p\bar p\to L^\pm\to
\mu\nu_\mu+X$. For all the cases
we have also applied a cut $\vert p_t^{miss}\vert_c=\vert p_t^\mu\vert_c$.
Here $\# B(S)$ corresponds to the number of background (signal) events.}
\end{table}

\begin{table}
\hfil
\vbox{\offinterlineskip
\halign{&#& \strut\quad#\hfil\quad\cr
\tableline
\tableline
&$g/g''$ && $M$ && $\Gamma_{L_3}$ && $\Gamma_{R_3}$ && 
$\vert p_t^\mu\vert_c$ && 
$m_{\mu^+\mu^-}$ &&  $\# B$ && $\# S$ && $S/{\sqrt{S+B}}$ & \cr
\tableline
&\omit&&GeV&&GeV&&GeV&&GeV&&GeV&&\omit&&\omit&&\omit& \cr
\tableline
&\omit&&\omit&&\omit&&\omit&&\omit&&\omit&&\omit&&\omit&&\omit&\cr
&0.12 && 400 && 0.4 && 0.06 && 150 && 300 && 269 && 257&& 11.2 & \cr  
&0.20 && 600 && 1.7 && 0.03 && 200 && 500 && 33 && 106 && 9.0 & \cr  
&0.40 && 1000 && 11.1 && 1.64 && 300 && 800 && 1 && 3 && 1.5 & \cr 
&\omit&&\omit&&\omit&&\omit&&\omit&&\omit&&\omit&&\omit&&\omit&\cr
\tableline
\tableline}}
\caption{Degenerate BESS at Tevatron Upgrade  for the process
$p\bar p\to L_3,R_3\to
\mu^+\mu^- +X$. Here $\# B(S)$ corresponds to the number of background 
(signal) events.}
\end{table}

\begin{table}
\hfil
\vbox{\offinterlineskip
\halign{&#& \strut\quad#\hfil\quad\cr
\tableline
\tableline
&$g/g''$ && $M$ && $\Gamma_{L^\pm}$ && $\vert p_t^\mu\vert_c$ && 
$m_T$ &&  $\# B$ && $\# S$ && $S/{\sqrt{S+B}}$ & \cr
\tableline
&\omit&&GeV&&GeV&&GeV&&GeV&&\omit&&\omit&&\omit& \cr
\tableline
&\omit&&\omit&&\omit&&\omit&&\omit&&\omit&&\omit&&\omit&\cr
&0.075 && 500 && 0.2 && 150 && 400 && 26300 && 20780 && 95.8 & \cr  
&0.15 && 500 && 0.8 && 150 && 400 && 85477 && 26300 && 255.0 & \cr  
&0.1 && 1000 && 0.7 && 300 && 800 && 2050 && 3130 && 43.5 & \cr
&0.1 && 1500 && 1.0 && 500 && 1300 && 247 && 469 && 17.5 & \cr  
&0.1 && 2000 && 1.4 && 700 && 1800 && 41 && 118 && 9.4 & \cr 
&\omit&&\omit&&\omit&&\omit&&\omit&&\omit&&\omit&&\omit&\cr
\tableline
\tableline}}
\caption{Degenerate BESS at LHC for the process $p p\to L^\pm\to
\mu\nu_\mu+X$. For all the cases
we have applied a cut $\vert p_t^{miss}\vert_c=\vert p_t^\mu\vert_c$.
 Here $\# B(S)$ corresponds to the number of background (signal) events.}
\end{table}

\begin{table}
\hfil
\vbox{\offinterlineskip
\halign{&#& \strut\quad#\hfil\quad\cr
\tableline
\tableline
&$g/g''$ && $M$ && $\Gamma_{L_3}$ && $\Gamma_{R_3}$ && 
$\vert p_t^\mu\vert_c$ && 
$m_{\mu^+\mu^-}$ &&  $\# B$ && $\# S$ && $S/{\sqrt{S+B}}$ & \cr
\tableline
&\omit&&GeV&&GeV&&GeV&&GeV&&GeV&&\omit&&\omit&&\omit& \cr
\tableline
&\omit&&\omit&&\omit&&\omit&&\omit&&\omit&&\omit&&\omit&&\omit&\cr
&0.075 && 500 && 0.2 && 0.03 && 150 && 400 && 16781 && 4300 && 29.6 & \cr 
&0.15 && 500 && 0.8 && 0.11 && 150 && 400 && 17480 && 16781 && 94.4 & \cr  
&0.10 && 1000 && 0.7 && 0.10 && 300 && 800 && 1145 && 605 && 14.5 & \cr  
&0.10 && 1500 && 1.0 && 0.15 && 500 && 1300 && 146 && 153 && 8.8 & \cr  
&0.10 && 2000 && 1.4 && 0.20 && 700 && 1800 && 35 && 22 && 2.9 & \cr  
&\omit&&\omit&&\omit&&\omit&&\omit&&\omit&&\omit&&\omit&&\omit&\cr
\tableline
\tableline}}
\caption{Degenerate BESS at LHC for the process $p p\to L_3,R_3\to
\mu^+\mu^- +X$. Here $\# B(S)$ corresponds to the number of background 
(signal) events.}
\end{table}

\begin{figure} 
\epsfxsize=8truecm
\centerline{\epsffile{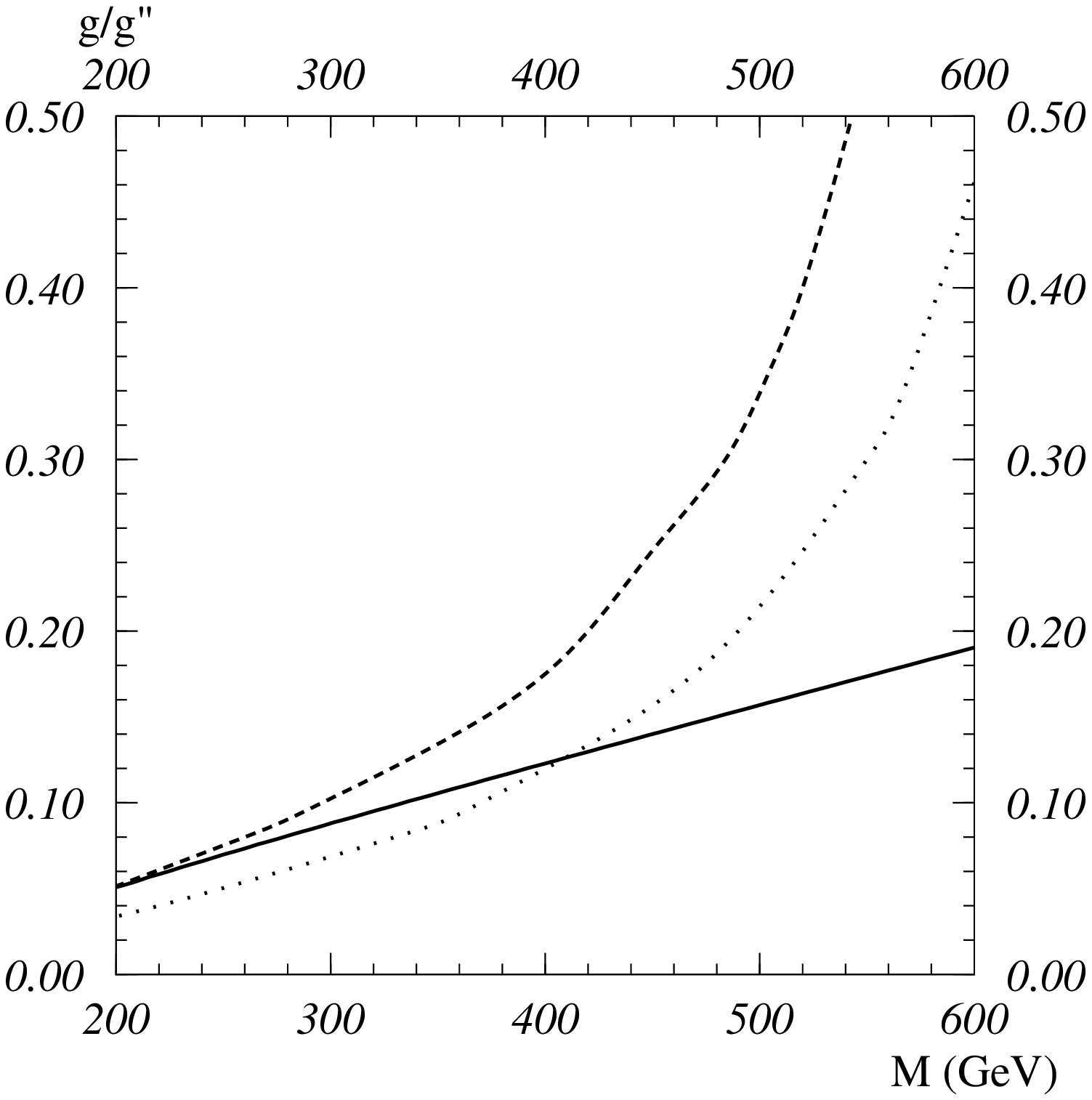}}
\noindent
{\bf Fig. 1} - { $ 95 \% $ C.L. upper bounds on $g/g''$ vs. M from 
LEPI data (solid line) and CDF with $L=19.7~pb^{-1}$ (dashed line).
The dotted line shows the extrapolation of the CDF bounds to $L=100~ pb^{-1}$.}
\end{figure} 

\begin{figure}
\epsfysize=8truecm
\centerline{\epsffile{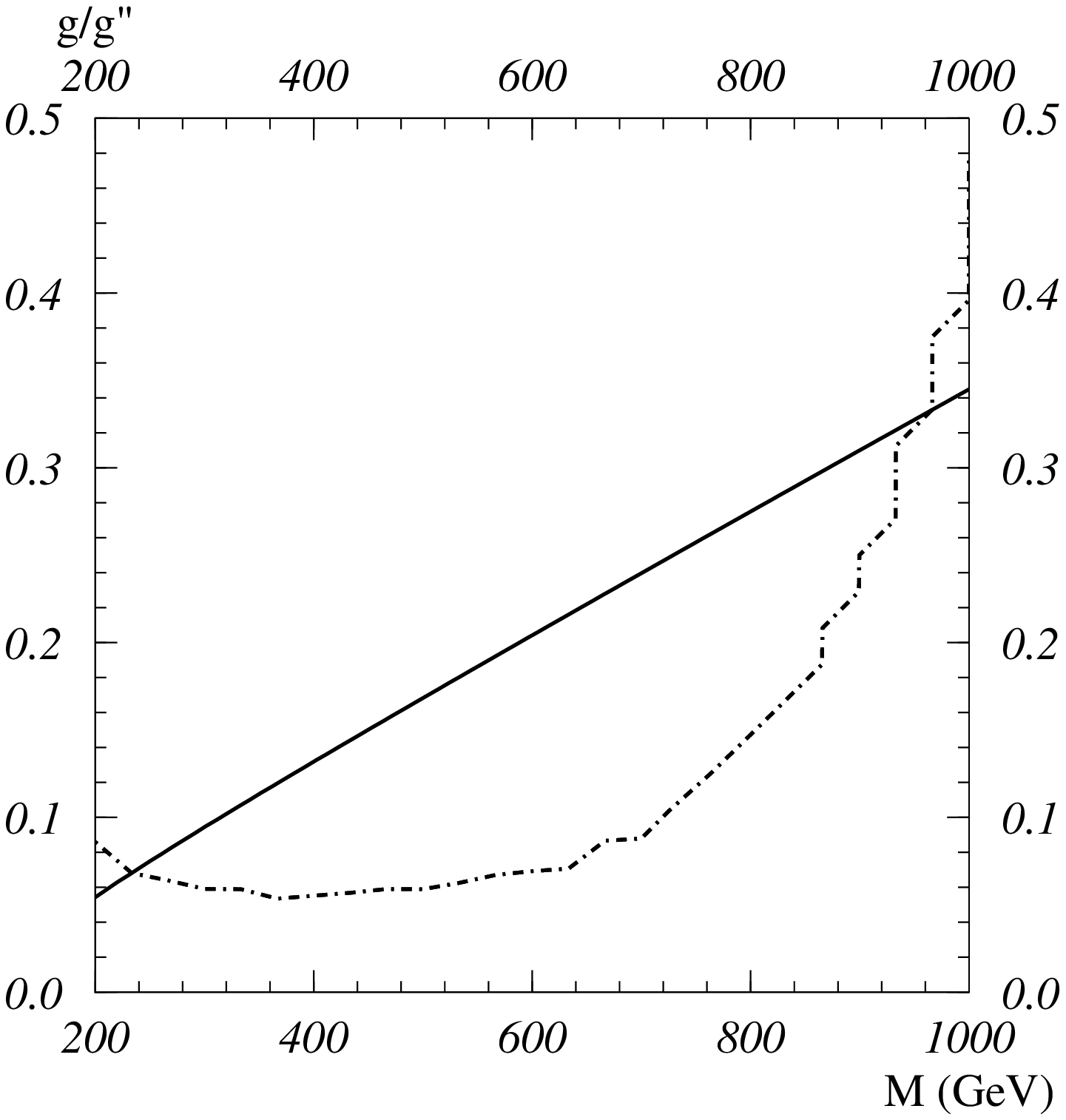}}
\noindent
{\bf Fig. 2} - {90\% C.L. limits on the parameter space $(M,g/g'')$ of 
degenerate BESS model at Tevatron upgrade with $\sqrt{s} =2$ TeV and a 
luminosity of $10^{33} cm^{-2} s^{-1}$, supposing no deviation is seen with 
respect to the SM in the total cross-section $p {\bar p} \to 
\mu \nu_\mu$. A minimum of 10 events per year is 
required to detect the signal with respect to the background; both the 
statistical error and a systematic error of 5\% on the cross-section 
are taken into account. The applied cuts are $\vert p_{T\mu}\vert >M/2 - 
50$ GeV. The figure is obtained from a grid of 25x25 cross-section 
points in the parameter space of the model. The continuous 
line corresponds to the LEPI limits.}
\end{figure}

\begin{figure}
\epsfysize=8truecm
\centerline{\epsffile{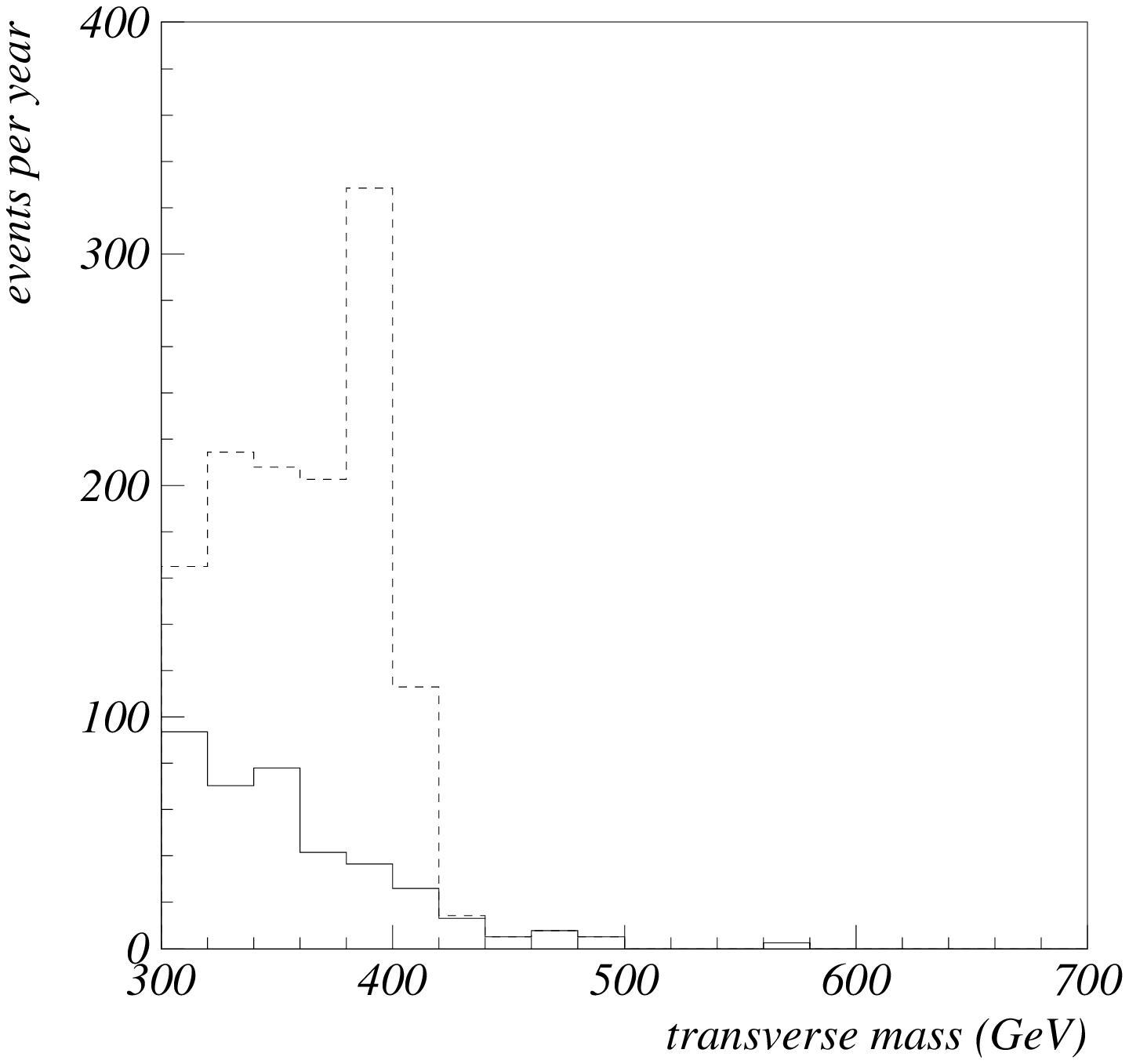}}
\noindent
{\bf Fig. 3} - {Transverse mass differential distribution of 
$p \bar p \to L^\pm, W^\pm \to \mu \nu_\mu$ events at Tevatron upgrade with 
a luminosity of $10^{33} cm^{-2} s^{-1}$ and $\sqrt{s} =2$ TeV, 
for $M=400$ GeV, $g/g''=0.12$.
The following cuts have been applied: $\vert p_{T\mu}\vert >150$ GeV,
$m_{T} >300$ GeV. The continuous line is the SM background, 
the dashed line represents the degenerate BESS model signal plus background. 
The number of signal events per year is 887, the corresponding background 
consists of 385 events.}
\end{figure}

\begin{figure}
\epsfysize=8truecm
\centerline{\epsffile{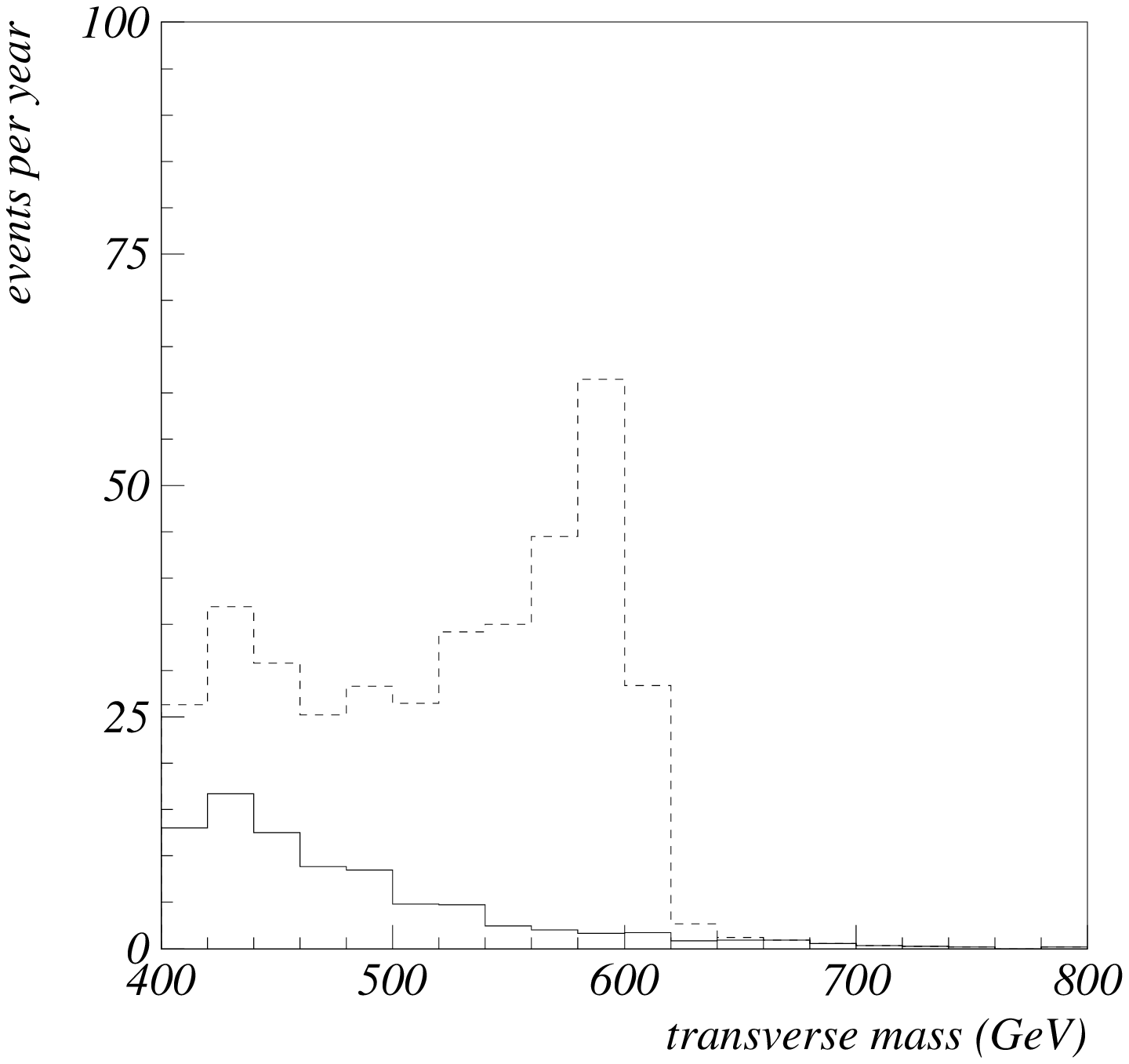}}
\noindent
{\bf Fig. 4} - {Same as in Fig. 3, 
for $M=600$ GeV, $g/g''=0.2$.
The following cuts have been applied: $\vert p_{T\mu}\vert >200$ GeV,
$m_{T} >400$ GeV.
The number of signal events per year is 303, the corresponding background 
consists of 82 events.}
\end{figure}

\begin{figure}
\epsfysize=8truecm
\centerline{\epsffile{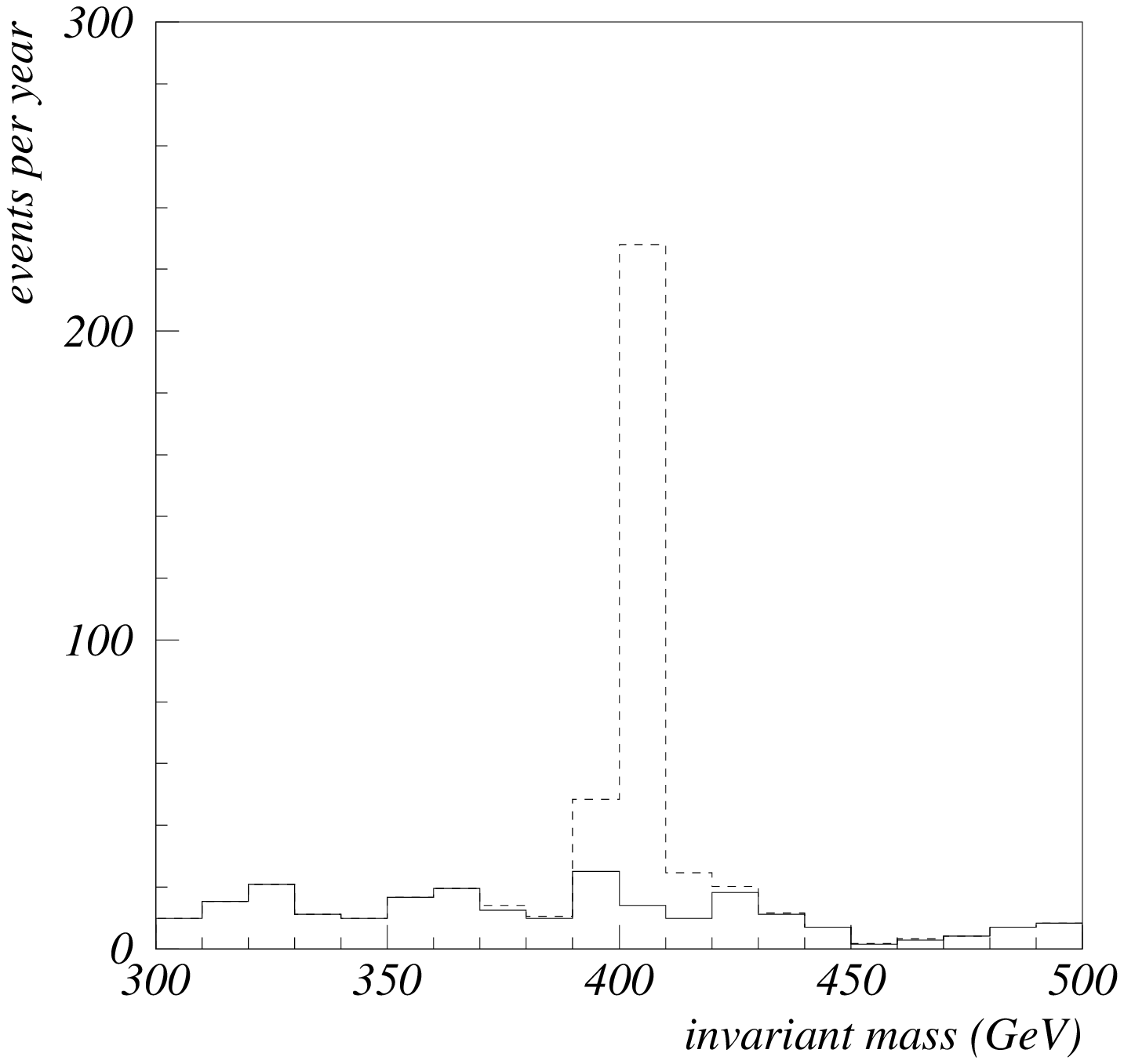}}
\noindent
{\bf Fig. 5} - {Invariant mass differential distribution of 
$p \bar p \to L_3, R_3, Z, \gamma \to \mu^+ \mu^-$ events at Tevatron upgrade with 
a luminosity of $10^{33} cm^{-2} s^{-1}$ and $\sqrt{s} =2$ TeV, 
for $M=400$ GeV, $g/g''=0.12$. The following cuts have been applied: 
$\vert p_{T\mu}\vert >150$ GeV, $m_{\mu^+ \mu^-} >300$ GeV. 
The continuous line is the SM background, 
the dashed line represents the degenerate BESS model signal plus background. 
The number of signal events per year is 257, the corresponding background 
consists of 269 events.}
\end{figure}

\begin{figure}
\epsfysize=8truecm
\centerline{\epsffile{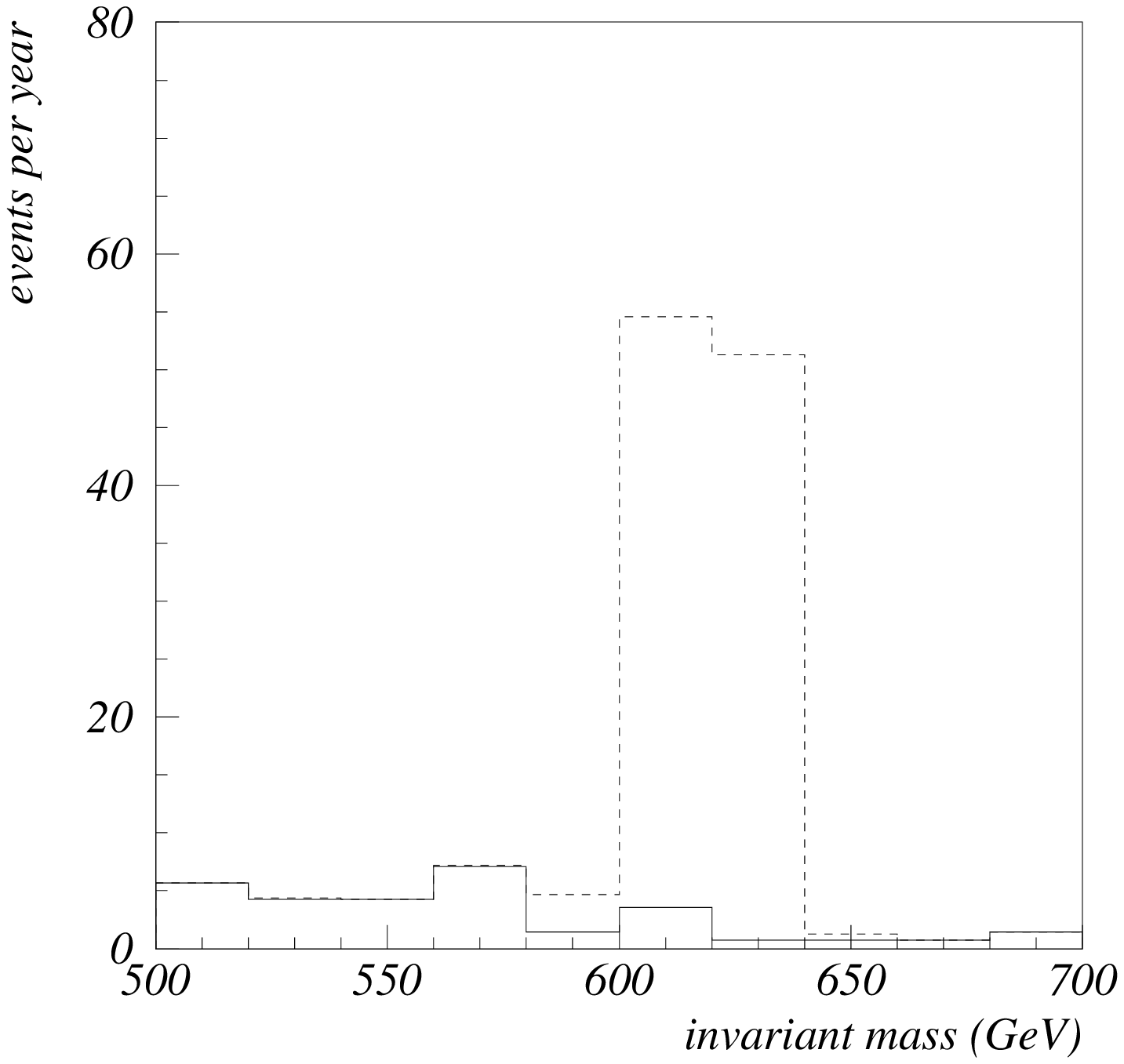}}
\noindent
{\bf Fig. 6} - {Same as in Fig. 5, 
for $M=600$ GeV, $g/g''=0.2$. The following cuts have been applied: 
$\vert p_{T\mu}\vert >200$ GeV, $m_{\mu^+ \mu^-} >500$ GeV. 
The number of signal events per year is 106, the corresponding background 
consists of 33 events.}
\end{figure}

\begin{figure}
\epsfysize=8truecm
\centerline{\epsffile{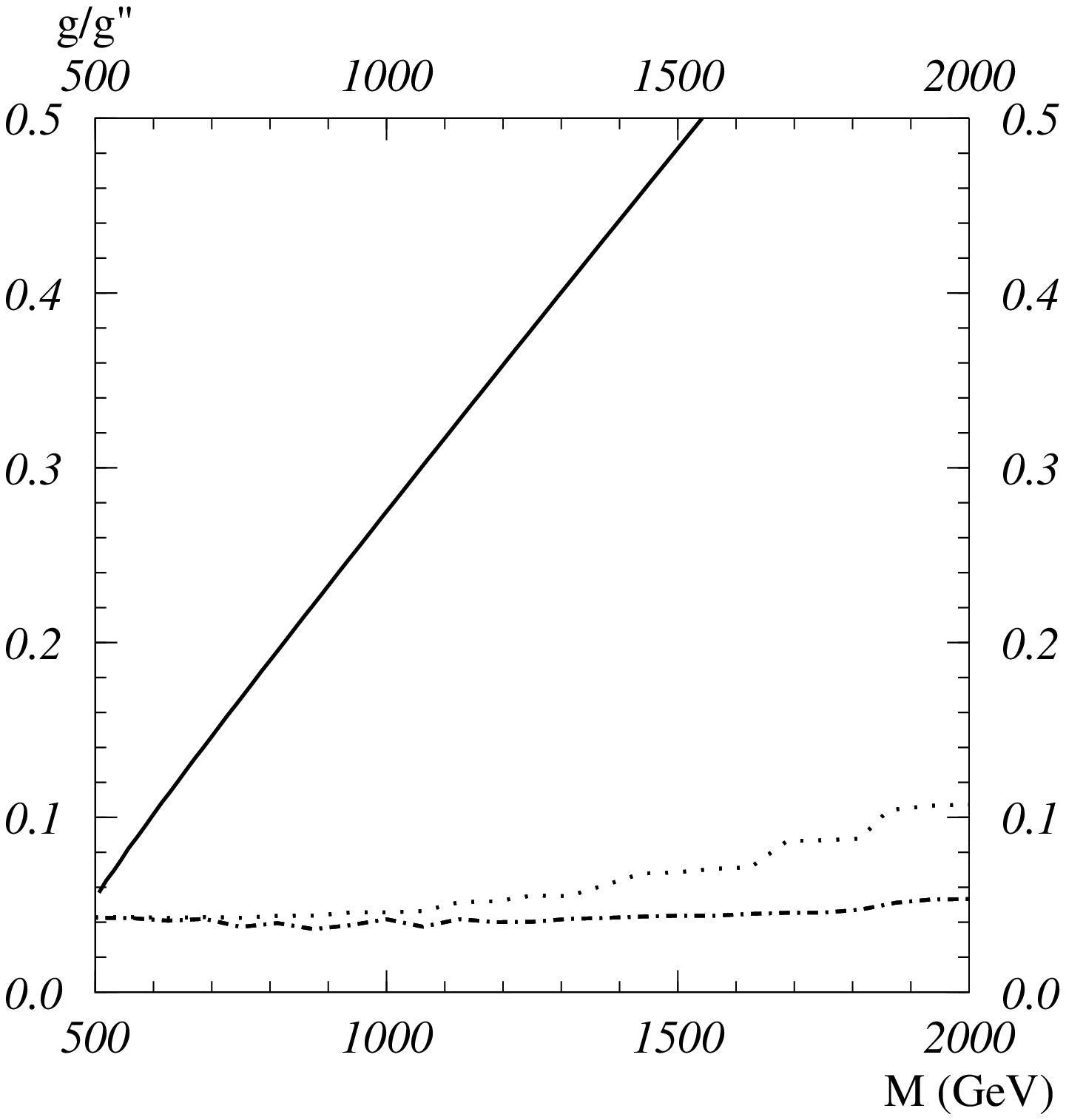}}
\noindent
{\bf Fig. 7} - {90\% C.L. limits on the parameter space $(M,g/g'')$ of 
degenerate 
BESS model at LHC with $\sqrt{s} =14$ TeV and a luminosity of $10^{34} cm^{-2} 
s^{-1}$ (dotted-dashed line) or $10^{33} cm^{-2} s^{-1}$ (dotted line), 
supposing no deviation is seen with respect to the SM in the 
total cross-section $pp \to \mu \nu_\mu$. A minimum of 10 events per year is 
required to detect the signal with respect to the background; both the 
statistical error and a systematic error of 5\% on the cross-section 
are taken into account. The applied cuts are $\vert p_{T\mu}\vert >M/2 - 
50$ GeV.  The figure is obtained from a grid of 25x25 cross-section 
points in the parameter space of the model. The continuous line 
corresponds to LEPI limits.}
\end{figure}

\begin{figure}
\epsfysize=8truecm
\centerline{\epsffile{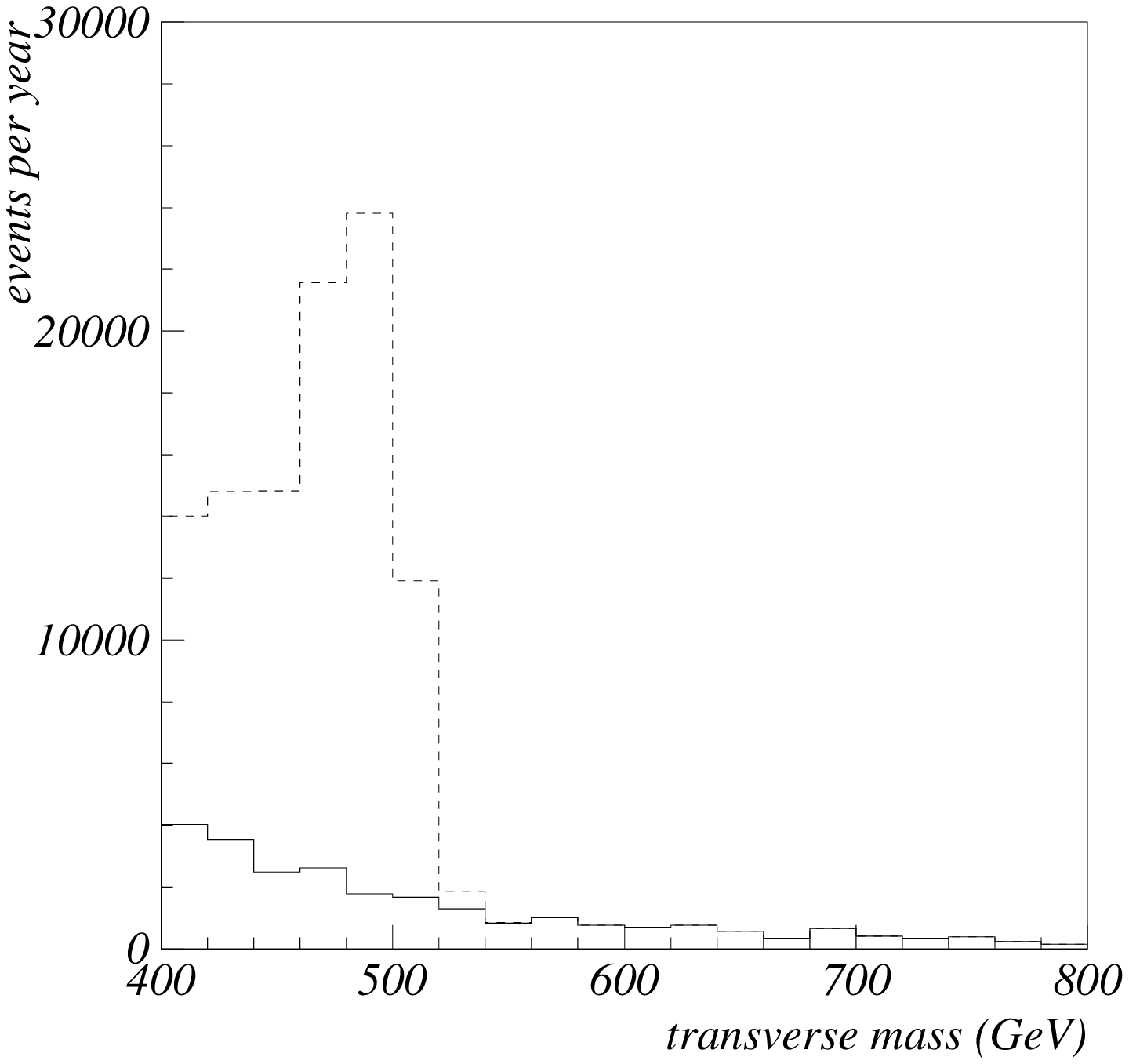}}
\noindent
{\bf Fig. 8} - {Transverse mass differential distribution of 
$pp \to L^\pm, W^\pm \to \mu \nu_\mu$ events at LHC with a luminosity of
$10^{34} cm^{-2} s^{-1}$ and $\sqrt{s} =14$ TeV, 
for $M=500$ GeV, $g/g''=0.15$.
The following cuts have been applied: $\vert p_{T\mu}\vert >150$ GeV,
$m_{T} >400$ GeV. The continuous line is the SM background, 
the dashed line represents the degenerate BESS model signal plus background. 
The number of signal events per year is 85477, the corresponding background 
consists of 26300 events.}
\end{figure}

\begin{figure}
\epsfysize=8truecm
\centerline{\epsffile{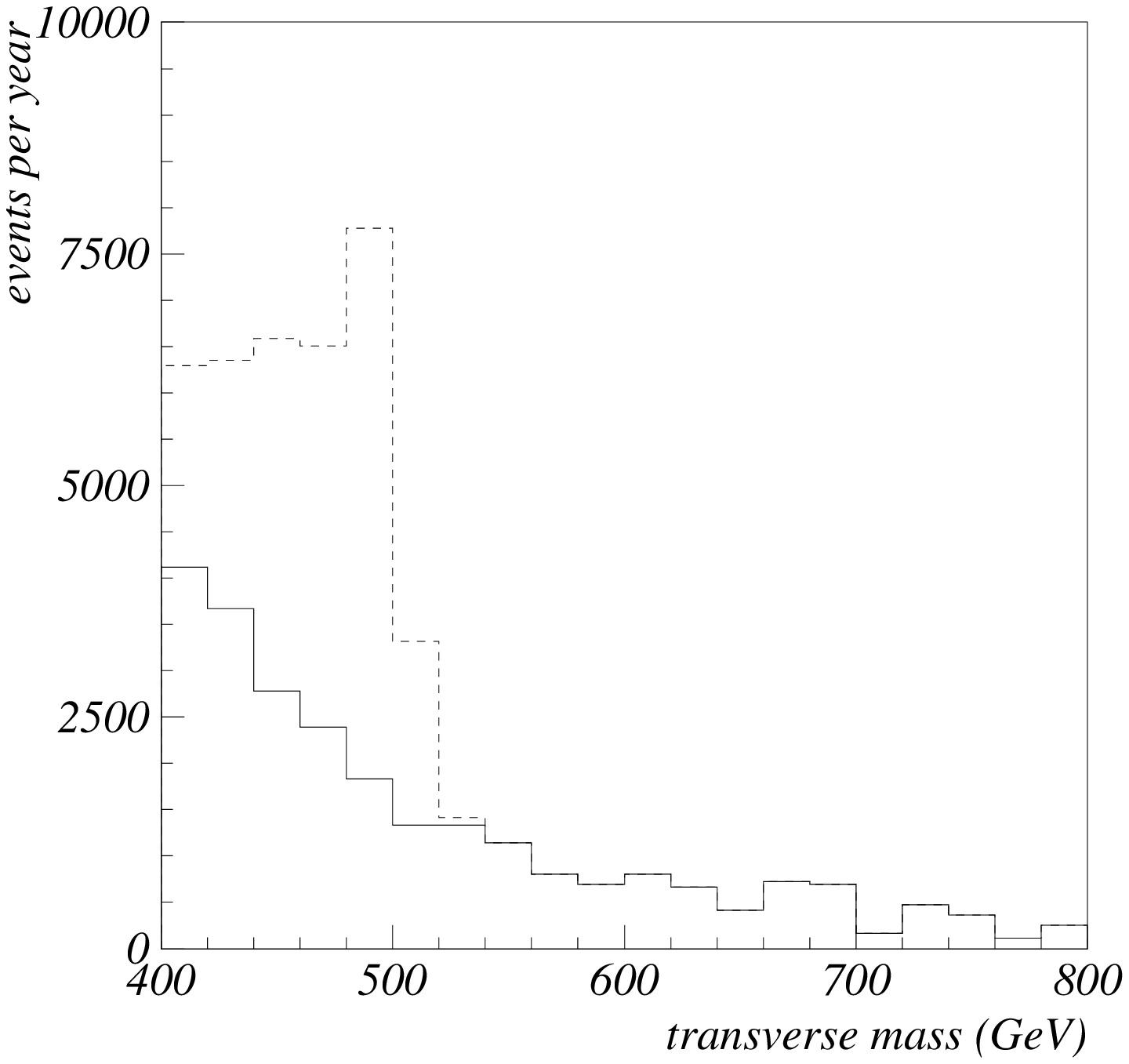}}
\noindent
{\bf Fig. 9} - {Same as in Fig. 8, 
for $M=500$ GeV, $g/g''=0.075$.
The following cuts have been applied: $\vert p_{T\mu}\vert >150$ GeV,
$m_{T} >400$ GeV.  
The number of signal events per year is 20780, the corresponding background 
consists of 26300 events.}
\end{figure}

\begin{figure}
\epsfysize=8truecm
\centerline{\epsffile{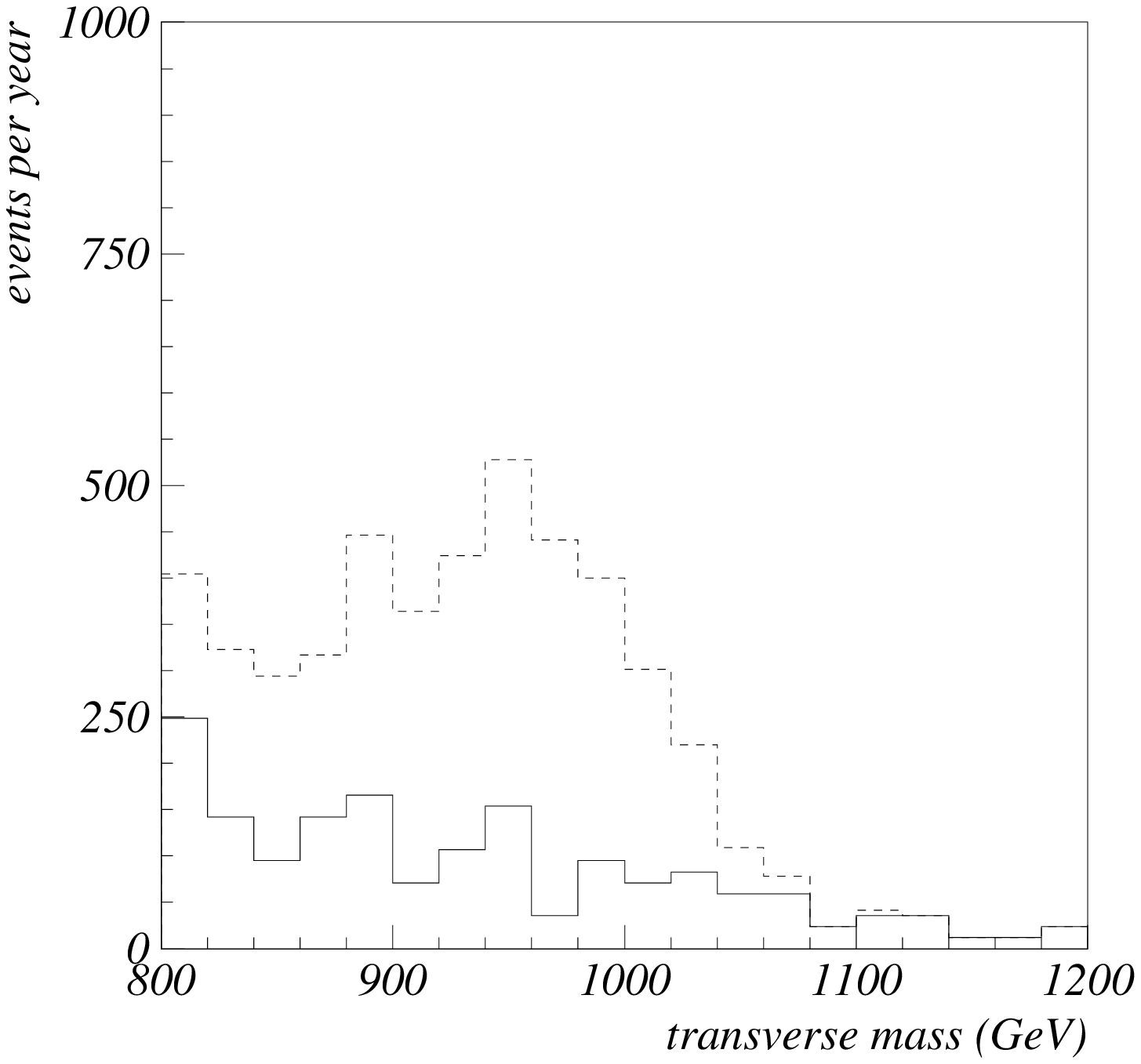}}
\noindent
{\bf Fig. 10} - {Same as in Fig. 8, 
for $M=1000$ GeV, $g/g''=0.1$.
The following cuts have been applied: $\vert p_{T\mu}\vert >300$ GeV,
$m_{T} >800$ GeV.
The number of signal events per year is 3130, the corresponding background 
consists of 2050 events.}
\end{figure}

\begin{figure}
\epsfysize=8truecm
\centerline{\epsffile{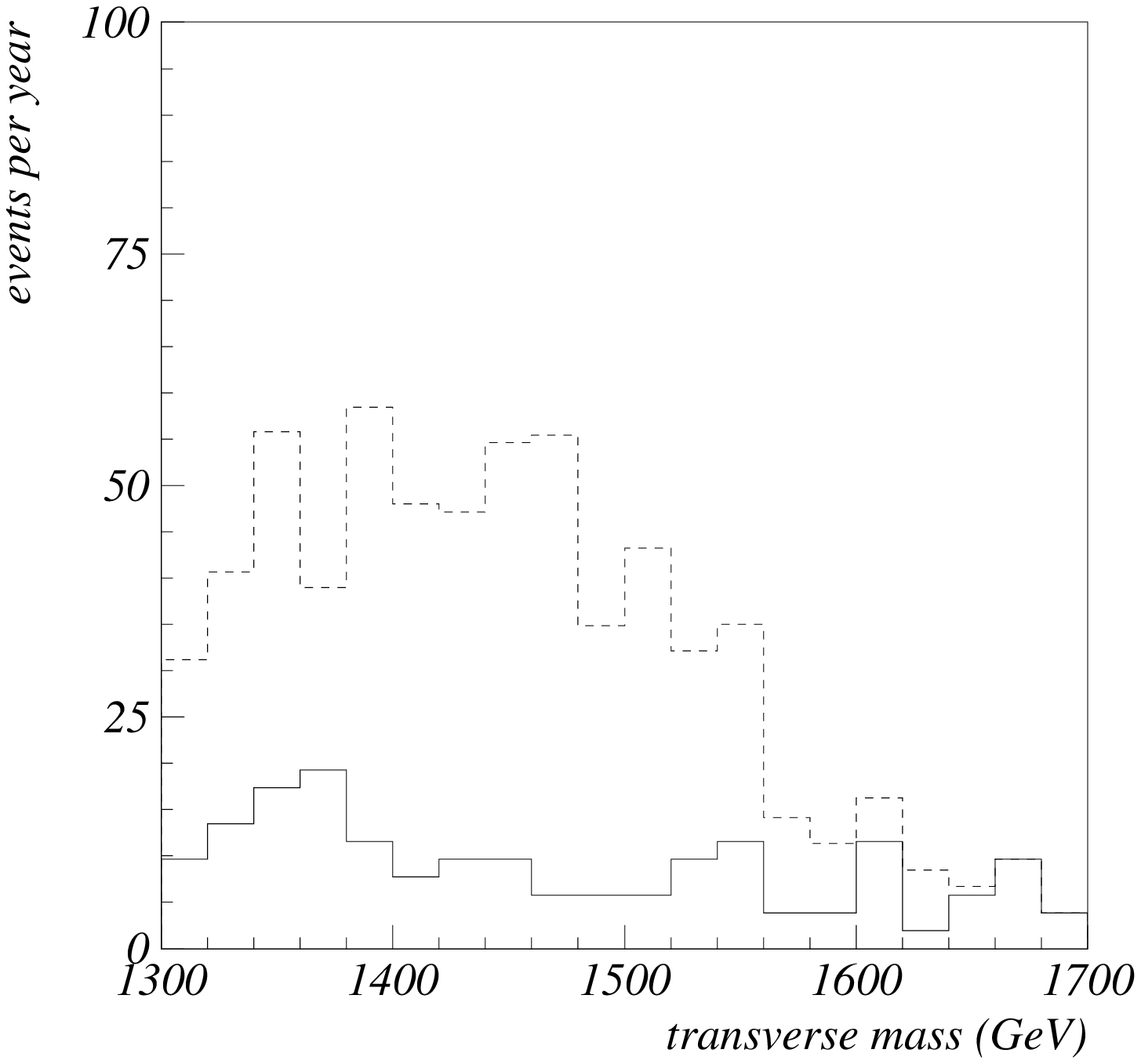}}
\noindent
{\bf Fig. 11} - {Same as in Fig. 8, 
for $M=1500$ GeV, $g/g''=0.1$.
The following cuts have been applied: $\vert p_{T\mu}\vert >500$ GeV,
$ m_{T} >1300 $ GeV. The number of signal events per year is 469, the 
corresponding background consists of 247 events.}
\end{figure}

\begin{figure}
\epsfysize=8truecm
\centerline{\epsffile{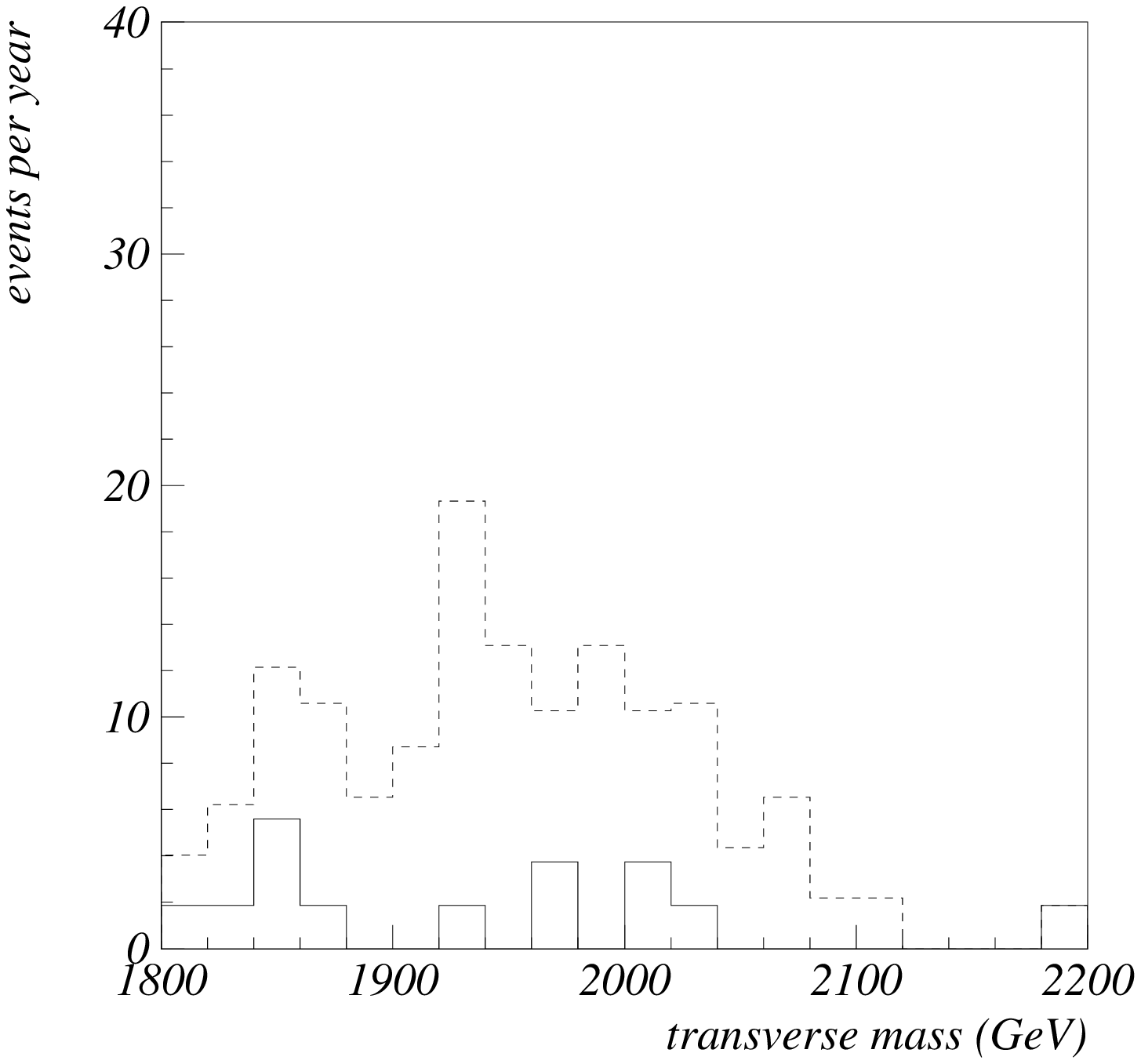}}
\noindent
{\bf Fig. 12} - {Same as in Fig. 8, 
for $M=2000$ GeV, $g/g''=0.1$.
The following cuts have been applied: $\vert p_{T\mu}\vert >700$ GeV,
$m_{T} >1800$ GeV.
The number of signal events per year is 118, the corresponding background 
consists of 41 events.}
\end{figure}

\begin{figure}
\epsfysize=8truecm
\centerline{\epsffile{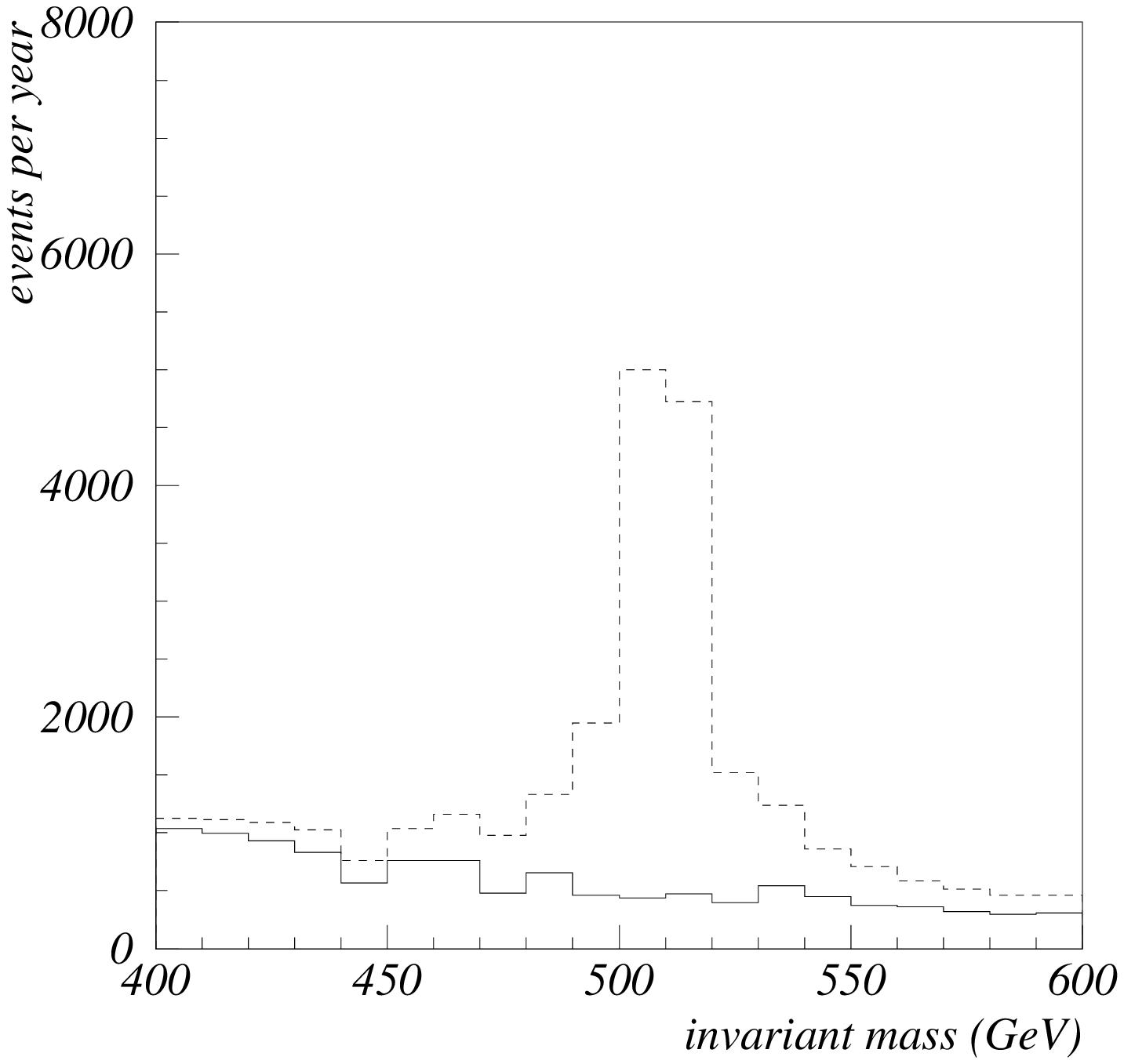}}
\noindent
{\bf Fig. 13} - {Invariant mass differential distribution of 
$pp \to L_3, R_3, Z, \gamma \to \mu^+ \mu^-$ events at LHC with 
a luminosity of $10^{34} cm^{-2} s^{-1}$ and $\sqrt{s} =14$ TeV, 
for $M=500$ GeV, $g/g''=0.15$. The following cuts have been applied: 
$\vert p_{T\mu}\vert >150$ GeV, $m_{\mu^+ \mu^-} >400$ GeV. 
The continuous line is the SM background, 
the dashed line represents the degenerate BESS model signal plus background. 
The number of signal events per year is 17480, the corresponding background 
consists of 16781 events.}
\end{figure}

\begin{figure}
\epsfysize=8truecm
\centerline{\epsffile{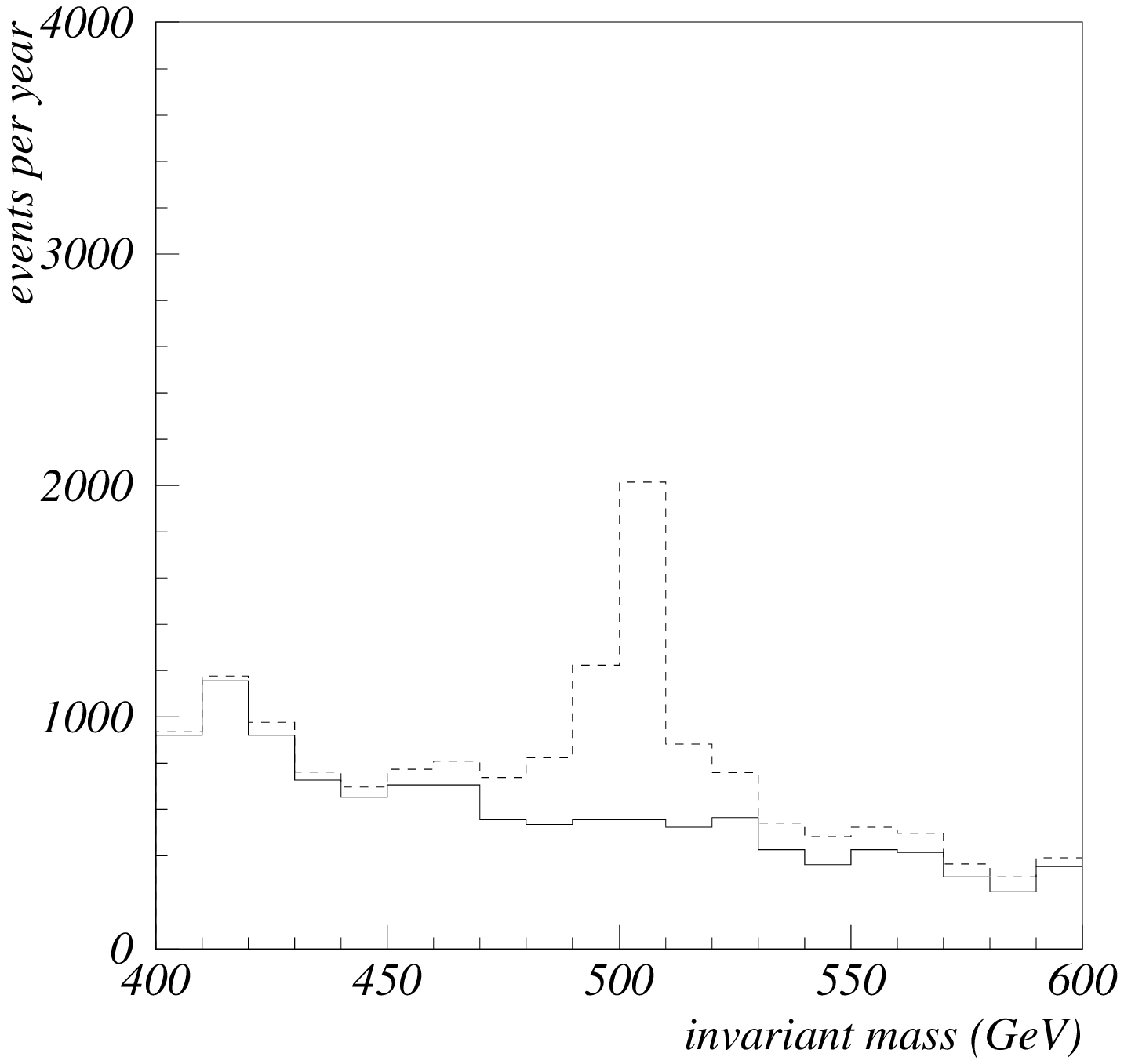}}
\noindent
{\bf Fig. 14} - {Same as in Fig. 13,
for $M=500$ GeV, $g/g''=0.075$. The following cuts have been applied: 
$\vert p_{T\mu}\vert >150$ GeV, $m_{\mu^+ \mu^-} >400$ GeV. 
The number of signal events per year is 4300, the corresponding background 
consists of 16781 events.}
\end{figure}

\begin{figure}
\epsfysize=8truecm
\centerline{\epsffile{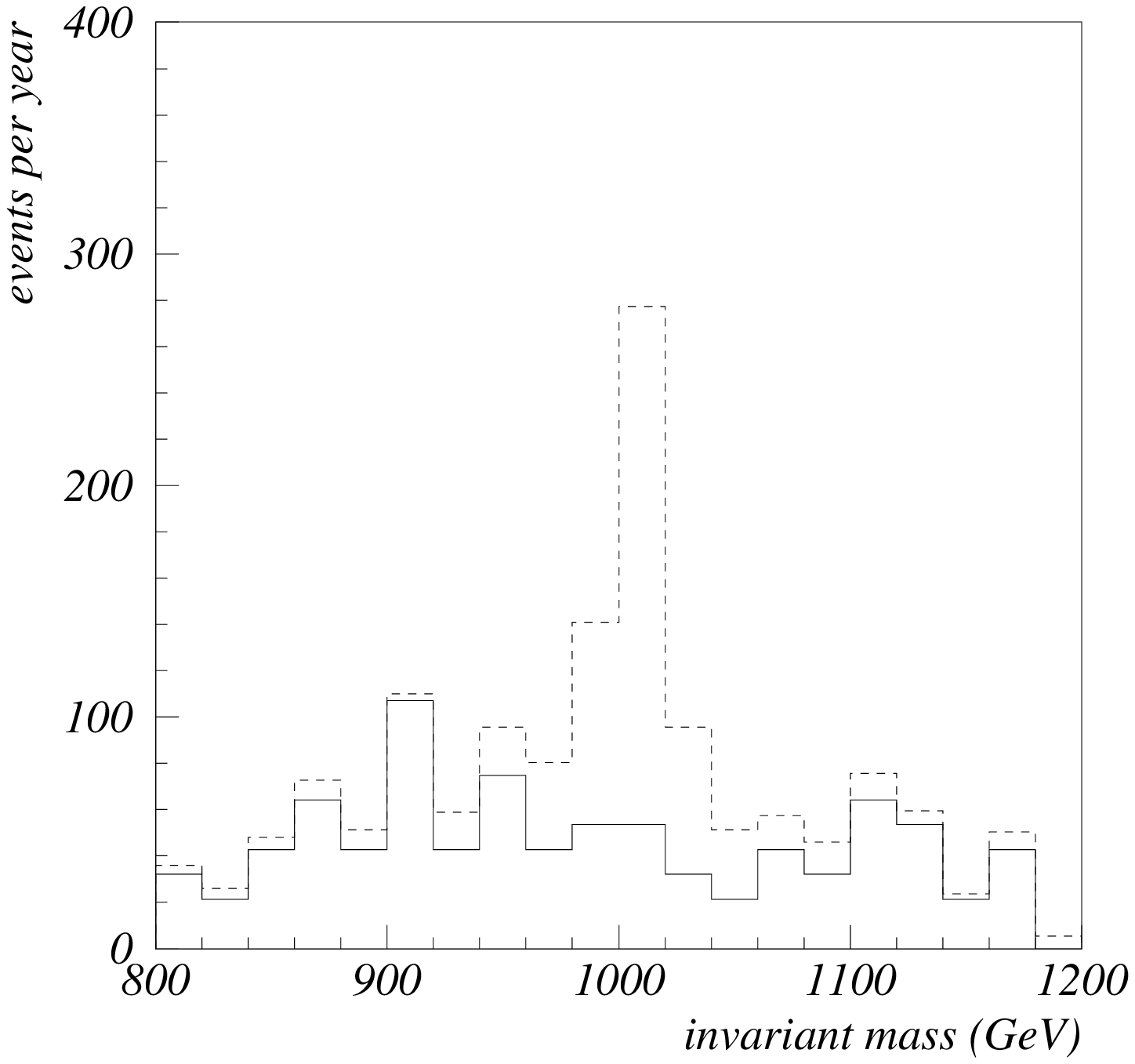}}
\noindent
{\bf Fig. 15} - {Same as in Fig. 13,
for $M=1000$ GeV, $g/g''=0.1$. The following cuts have been applied: 
$\vert p_{T\mu}\vert >300$ GeV, $m_{\mu^+ \mu^-} >800$ GeV. 
The number of signal events per year is 605, the corresponding background 
consists of 1145 events.}
\end{figure}

\begin{figure}
\epsfysize=8truecm
\centerline{\epsffile{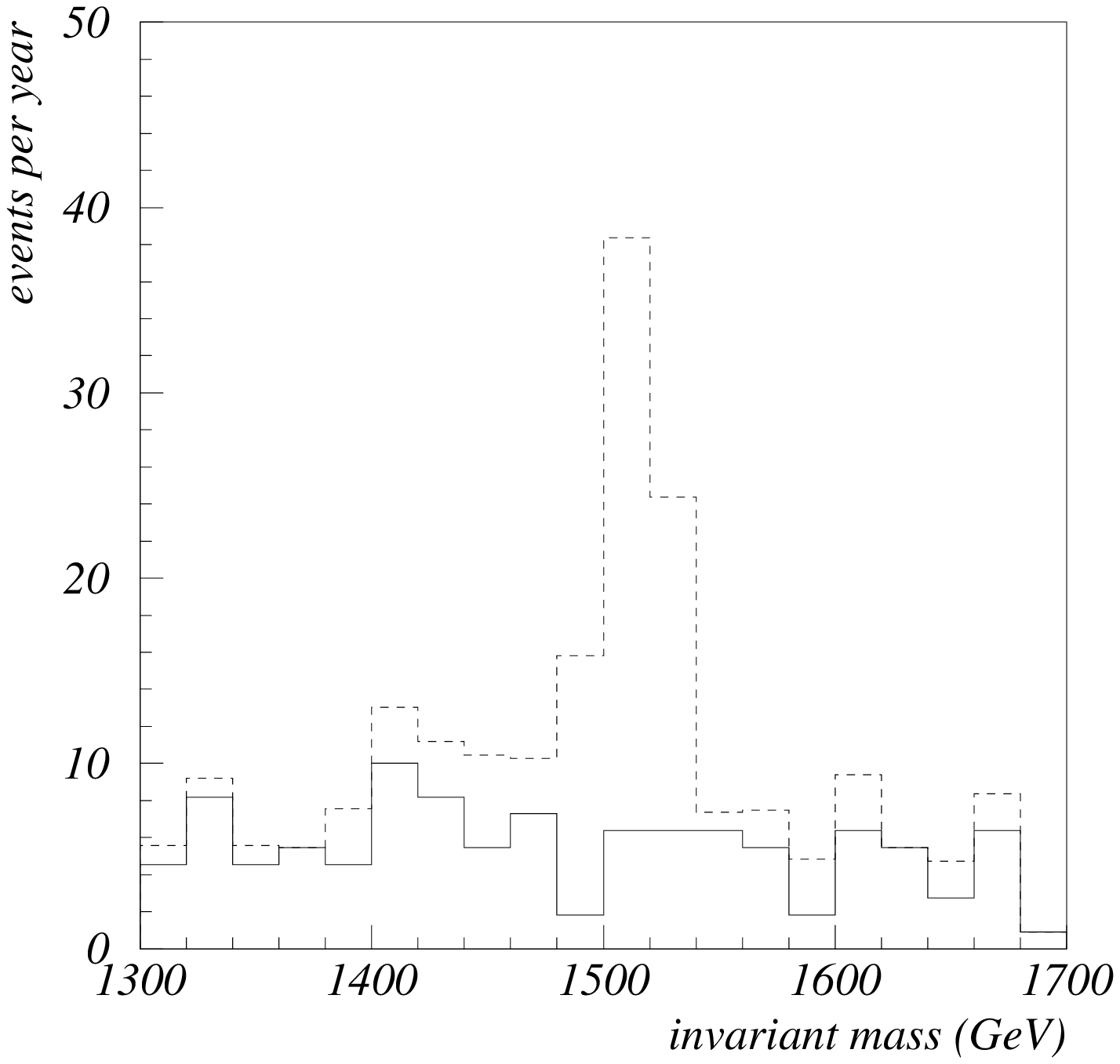}}
\noindent
{\bf Fig. 16} - {Same as in Fig. 13,
for $M=1500$ GeV, $g/g''=0.1$. The following cuts have been applied: 
$\vert p_{T\mu}\vert >500$ GeV, $m_{\mu^+ \mu^-} >1300$ GeV. 
The number of signal events per year is 153, the corresponding background 
consists of 146 events.}
\end{figure}

\begin{figure}
\epsfysize=8truecm
\centerline{\epsffile{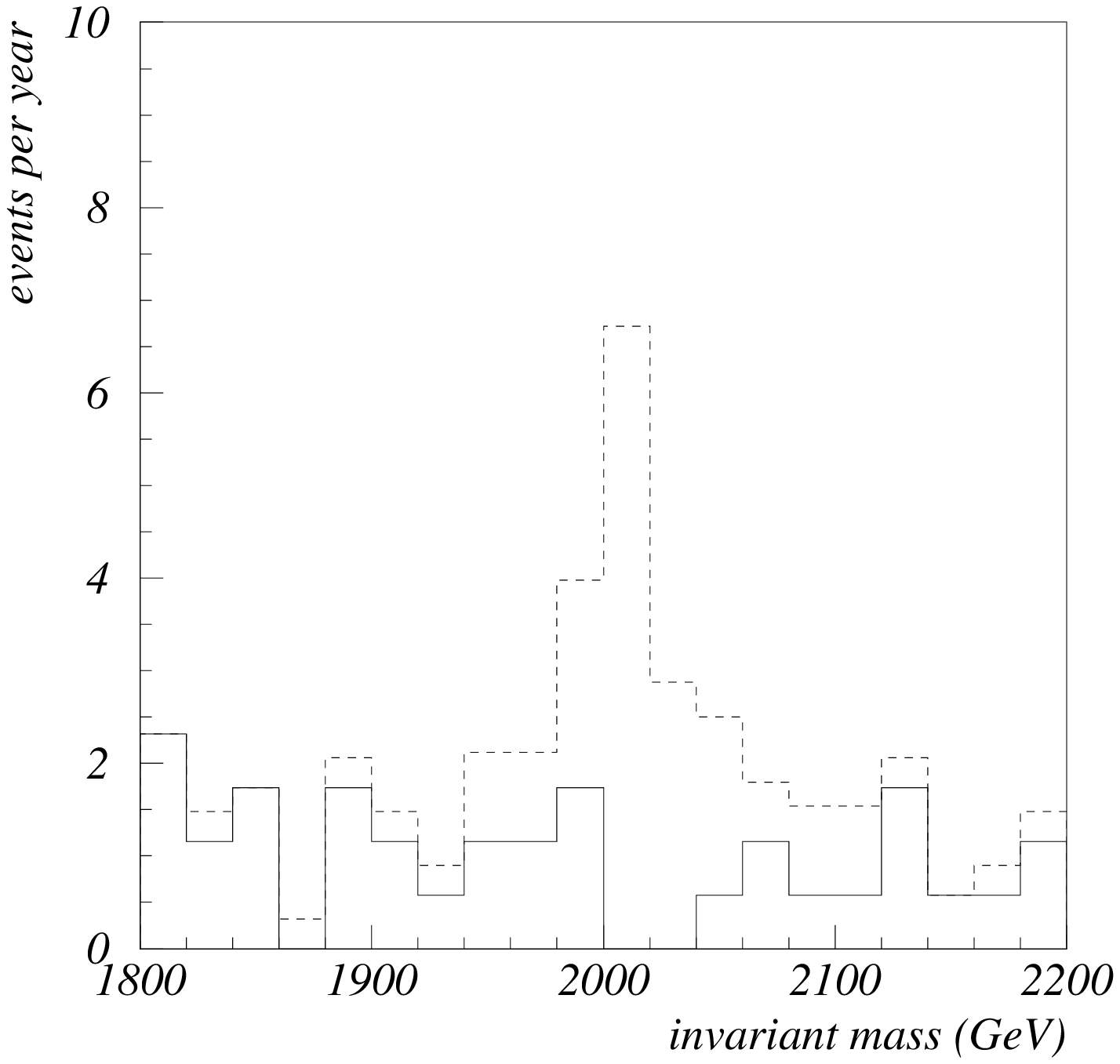}}
\noindent
{\bf Fig. 17} - {Same as in Fig. 13,
for $M=2000$ GeV, $g/g''=0.1$. The following cuts have been applied: 
$\vert p_{T\mu}\vert >700$ GeV, $m_{\mu^+ \mu^-} >1800$ GeV. 
The number of signal events per year is 22, the corresponding background 
consists of 35 events.}
\end{figure}

\end{document}